\documentclass[11pt,a4paper]{article}%
\pdfoutput=1

\usepackage[english]{babel}
\usepackage[nosort,compress]{cite}
\usepackage{graphicx,epsfig,subfigure}
\usepackage[dvipsnames]{xcolor}
\usepackage[linktoc=page,bookmarks=false,colorlinks=false,
linkbordercolor=RoyalBlue,citebordercolor=ForestGreen,urlbordercolor=CornflowerBlue]{hyperref}
\usepackage[normal,font=small,labelfont=bf,labelsep=period]{caption}
\usepackage{amsmath,amssymb,amsfonts,array,enumerate,slashed,multirow,fancybox,marvosym,wasysym,soul}
\usepackage[normalem]{ulem}

\usepackage[margin=1in]{geometry}
\font\tenrsfs=rsfs10 at 12pt

\font\sevenrsfs=rsfs7
\font\fiversfs=rsfs5
\newfam\rsfsfam
\textfont\rsfsfam=\tenrsfs
\scriptfont\rsfsfam=\sevenrsfs
\scriptscriptfont\rsfsfam=\fiversfs
\def\mathscr#1{{\fam\rsfsfam\relax#1}}

\font\teneusm=eusm10 at 12pt
\newfam\eusmfam
\textfont\eusmfam=\teneusm

\font\tenbbm=bbm10 at 12pt
\newfam\bbmfam
\textfont\bbmfam=\tenbbm

\def\L{\mathcal{L}}

\def\B{\mathcal{B}}
\def\BR{\mathcal{B}}

\renewcommand{\Im}{\mathop{\rm Im}}

\def\BK{\B(K^+\to \pi^+\nu\bar\nu)}
\def\BB{\B(B\to K^{(*)}\nu\bar\nu)}
\def\RD{R_{D^{(*)}}}

\numberwithin{equation}{section}
\title{\vspace{-2cm}\begin{flushright}
{\small ZU-TH-15/17}
\end{flushright}
\vspace{1cm}
\bf\LARGE Probing Lepton Flavour Universality \\
with \boldmath $K\to\pi\nu\bar\nu\unboldmath$ decays}
\author{\large Marzia Bordone, Dario Buttazzo, Gino Isidori, Joachim Monnard}
\date{\small {\it Physik-Institut, Universit\"at Z\"urich, Winterthurerstrasse 190, 8057 Z\"urich}}

\begin{document}

\begin{titlepage}
\maketitle

\thispagestyle{empty}

\begin{abstract}
\noindent
We analyse the rare processes $K\to\pi\nu\bar\nu$ in view of the recent hints of violations of Lepton Flavour Universality (LFU) observed in $B$ meson decays. If, as suggested by present data, the new interactions responsible for LFU violations couple mainly to the third generation of left-handed fermions, $K\to\pi\nu\bar\nu$ decays turn out to be particularly interesting: these are the only kaon decays with third-generation leptons (the  $\tau$ neutrinos) in the final state. In order to relate $B$-physics anomalies and $K$ decays we adopt an Effective Field Theory approach, assuming that the new interactions satisfy an approximate $U(2)_q\times U(2)_\ell$ flavour symmetry. In this framework we show that $O(1)$ deviations from the Standard Model predictions in $K\to\pi\nu\bar\nu$ branching ratios, closely correlated to similar effects in $B\to K^{(*)}\nu\bar\nu$, are naturally expected. The correlation of  $\mathcal{B}(K \to\pi \nu\bar\nu)$,  $\mathcal{B}(B\to K^{(*)}\nu\bar\nu)$, and the LFU violations in $B$ decays  would provide a very valuable tool to shed more light on this interesting phenomenon.
\end{abstract}

\end{titlepage}

 %%%%%%%%%%%%%%%%%%%%%%%%%%%%%%%%%%%%%%%%%%%%%%%%%%%%%%%%%%%
\section{Introduction}\label{sec:intro}
%%%%%%%%%%%%%%%%%%%%%%%%%%%%%%%%%%%%%%%%%%%%%%%%%%%%%%%%%%%

The hints of Lepton Flavour Universality (LFU) violations in semi-leptonic $B$ decays
are among the most interesting  deviations from the Standard Model (SM)
reported by experiments in the last few years. 
The very recent result on the $R_{K^*}$ ratio by LHCb~\cite{Aaij:2017vbb}
is only the last piece of a seemingly coherent set of results, involving different 
observables and different experiments, that started in 2003 with the BaBar 
results on $R_{D^{(*)}}$~\cite{Lees:2013uzd}.
The evidences collected so far can naturally be grouped into two categories,
according to the underlying quark-level transition:
\begin{itemize}
\item{} deviations from 
$\tau/(\mu,e)$ universality in charged currents of the type $b\to c \ell \bar\nu$
(observed in $B \to D^* \ell \overline{\nu}$ and  $B \to D \ell \overline{\nu}$  
decays~\cite{Lees:2013uzd,Hirose:2016wfn,Aaij:2015yra});
\item{} deviations from 
$\mu/e$ universality in neutral currents of the type $b\to s \ell \overline{\ell}$
(observed in $B \to K^* \ell \overline{\ell}$ and  $B \to K \ell \overline{\ell}$  
decays~\cite{Aaij:2014ora,Aaij:2017vbb}).
\end{itemize}
In both cases the combination of the results leads to an evidence, exceeding the $3\sigma$ level,
for LFU contributions  of non-SM origin, whose size is $O(10\%)$ compared to the corresponding 
charged- or neutral-current SM amplitudes.\footnote{Updated fits for the Wilson coefficients of the relevant 
low-energy effective Hamiltonians can be found e.g. in 
Ref.~\cite{Altmannshofer:2017yso,Capdevila:2017bsm,Geng:2017svp,Ciuchini:2017mik,DAmico:2017mtc} for $b\to s\ell\bar\ell$ 
and Ref.~\cite{Bernlochner:2017jka} for $b\to c\ell\bar\nu$.}

These deviations from the SM have triggered a series of theoretical speculations
about possible New Physics (NP) interpretations.  In particular, attempts to provide 
a combined/coherent explanation for  both charged- and neutral-current anomalies have been
presented  in Ref.~\cite{Bhattacharya:2014wla, Alonso:2015sja, Greljo:2015mma, Calibbi:2015kma,Bauer:2015knc,
Fajfer:2015ycq, Barbieri:2015yvd, Das:2016vkr, Boucenna:2016qad,Becirevic:2016yqi,  Hiller:2016kry, Bhattacharya:2016mcc,
Buttazzo:2016kid,Barbieri:2016las,Bordone:2017anc,Crivellin:2017zlb}.
 One of the puzzling aspects of present anomalies is that they have been seen only in semileptonic 
$B$ decays and are quite large compared to the corresponding SM  amplitudes. 
On the contrary, no evidences of deviations from the SM have been seen so far in the precise (per-mil) tests of LFU
performed in semileptonic $K$ and $\pi$ decays, in purely leptonic 
$\tau$ decays, and in electroweak precision observables. 
The most natural assumption to address this apparent paradox is the hypothesis 
that the NP responsible for the breaking of LFU is coupled mainly to the 
third generation of quarks and leptons, with some small (but non-negligible) mixing
with the light generations~\cite{Greljo:2015mma,Bordone:2017anc,Glashow:2014iga}. Within this paradigm, a  
motivated class of models are those based on a $U(2)_q\times U(2)_\ell$
 flavour symmetry acting 
on the light generations of SM fermions~\cite{Barbieri:2011ci,Barbieri:2012uh}, that turns 
out to be quite successful in addressing these 
anomalies while satisfying all existing bounds~\cite{Bordone:2017anc}.

If NP is coupled mainly to third generation fermions, it is very difficult to detect it in $K$ decays, which 
necessarily imply a transition among light quarks and, in most cases, 
also imply light leptons in the final states.  The only exception in this respect is 
provided by $K\to \pi\nu\bar{\nu}$ decays,  which involve third-generation leptons in the final state -- the $\tau$ neutrinos.
As we will show in the following, this fact implies that $K\to \pi\nu\bar{\nu}$ decays are a very sensitive 
probe of the most motivated models addressing  the hints of LFU violations in $B$ physics. 
On the one hand, $\BR(K\to \pi\nu\bar{\nu})$ could exhibit $O(1)$ deviations from the SM predictions 
in a large area of the parameter space of such models. On the other hand,  even in absence of large deviations,
improved measurements (or constraints) on  $\BR(K \to \pi\nu\bar{\nu})$ would provide 
a very valuable model-building information.

The paper is organised as follows: in Section~\ref{sec:kpnn} we briefly review the main formulae to 
evaluate $\BR(K\to\pi\nu\bar\nu)$ within and beyond the SM. In Section~\ref{sec:EFT} we
discuss the Effective Field Theory (EFT)  approach to LFU violations based  on the $U(2)_q\times U(2)_\ell$
flavour symmetry and, in that framework, we analyse the possible impact on 
$K\to\pi\nu\bar\nu$ decays. In Section~\ref{sec:obs} we focus in particular on the expected 
correlations between $K\to\pi\nu\bar\nu$, the $R_{D^{(*)}}$ anomaly, and $B\to K^{(*)}\nu\bar\nu$,
which turn out to be closely related observables (impact and constraints from other observables 
are briefly mentioned at the end of the section). 
The results are summarised in the Conclusions.

%%%%%%%%%%%%%%%%%%%%%%%%%%%%%%%%%%%%%%%%%%%%%%%%%%%%%%%%%%%
\boldmath
\section{The $K\to\pi\nu\bar\nu$ decays}\label{sec:kpnn}
\unboldmath
%%%%%%%%%%%%%%%%%%%%%%%%%%%%%%%%%%%%%%%%%%%%%%%%%%%%%%%%%%%

Here we briefly summarise the main steps to predict 
$\BR(K^+\to\pi^+\nu\bar\nu)$ and $\BR(K_L\to \pi^0\nu\bar\nu)$ within and beyond the SM, taking into account 
possible violations of LFU. 
The effective Lagrangian describing short-distance FCNC interactions of the type $d_L^i\to d_L^j\nu\bar\nu$ is 
\begin{equation}
\L_{\rm eff} = \frac{4 G_F}{\sqrt{2}}\frac{\alpha}{2\pi} V_{ti}^*V_{tj} C_{ij,\ell} \left(\bar d_L^i\gamma_\mu d_L^j\right)\left(\bar\nu_{\ell}\gamma^\mu \nu_{\ell}\right),\label{Leff}
\end{equation}
where $\alpha$ is the fine-structure constant, and $V_{ij}$ are the elements of the CKM matrix. 
For $s_L\to d_L\nu_\ell\bar\nu_\ell$, the Wilson coefficient in the SM reads
\begin{equation}
C^{{\rm SM}}_{sd,\ell} = -\frac{1}{s_w^2}\left( X_t + \frac{V_{cs}^*V_{cd}}{V_{ts}^*V_{td}} X_c^\ell\right),\label{CSM}
\end{equation}
where $X_t$ and  $X_c^\ell$ are the loop functions for the top and charm contributions, respectively, and $s_w$ is the sine of the weak mixing angle.

The branching ratio for $K^+\to \pi^+\nu\bar\nu$ in the SM, summing over the three neutrino species, can be written as \cite{Buchalla:1993wq}
\begin{equation}
\B(K^+\to\pi^+\nu\bar\nu)_{\rm SM} = \frac{\kappa_+ (1+\Delta_{\rm em})}{3} \sum_{\ell = e,\mu,\tau} \left| \frac{V_{ts}^*V_{td}}{\lambda^5}X_t + \frac{V_{cs}^*V_{cd}}{\lambda}\left(\frac{X_c^\ell}{\lambda^4} + \delta P_c^\ell\right)\right|^2,\label{Kpvv}
\end{equation}
where $\lambda$ is the Cabibbo angle, $\kappa_+ = (5.173\pm 0.025)\times 10^{-11}(\lambda/0.225)^8$, $\Delta_{\rm em} = -0.003$ is a QED correction~\cite{Mescia:2007kn},
and $\delta P_{c,u}^\ell \approx 0.04\pm 0.02$ is the long-distance contribution from light quark loops~\cite{Isidori:2005xm}. 
The numerical value of the loop functions are 
$X_t = 1.481\pm 0.009$ and $P_c = \frac{1}{3}\sum_\ell X_c^\ell/\lambda^4 = 0.365 \pm 0.012$~\cite{Buchalla:1998ba}.\footnote{The NLO
values of the individual $X_c^\ell$ can be found e.g.\ in \cite{Buras:2006gb}.}

Within the SM the CP-violating decay $K_L\to \pi^0\nu\bar\nu$ is lepton-flavour universal. However, in order to take into account possible 
violation of LFU beyond the SM, we can conveniently  write its branching ratio as
\begin{equation}
\B(K_L \to \pi^0\nu\bar\nu)_{\rm SM} =  \frac{\kappa_L}{3}  \sum_{\ell = e,\mu,\tau} 
\Im\left(\frac{V_{ts}^* V_{td}}{\lambda^5}X_t\right)^2,\label{KLvv}
\end{equation}
where $\kappa_L = (2.231\pm 0.013)\times 10^{-10} ( \lambda/ 0.225 )^8$.

In the class of NP models we will consider, the short-distance contributions to $K \to \pi \nu\bar\nu$ 
amplitudes are still left-handed but lepton flavour non-universal. 
The general expressions for the branching ratios  in presence of such non-standard contributions 
can simply be obtained replacing the function $X_t$ in \eqref{Kpvv} and \eqref{KLvv} by
\begin{align}
X(C_{sd,\ell}^{{\rm NP}}) = X_t + C_{sd,\ell}^{{\rm NP}}\, s^2_w,\label{XNP}
\end{align}
where $C^{\rm NP}_{sd,\ell}$ is the new physics contribution to the Wilson coefficient in \eqref{Leff}. 

Using the most recent determinations of the input parameters, the SM predictions for the two branching ratios are~\cite{Buras:2015qea}
\begin{align}
\B(K^+\to \pi^+\nu\bar\nu)_{\rm SM} &= (8.4\pm 1.0)\times 10^{-11},\label{KpSM}\\
\B(K_L\to \pi^0\nu\bar\nu)_{\rm SM} &= (3.4\pm 0.6)\times 10^{-11}\label{KLSM}.
\end{align}
The dominant source of error in \eqref{KpSM} and \eqref{KLSM} comes from the uncertainty in the CKM matrix elements, and from the charm contribution.

The current experimental bounds are~\cite{Olive:2016xmw}
\begin{align}
\B(K^+\to \pi^+\nu\bar\nu)_{\rm exp} &= 17.3^{+11.5}_{-10.5}\times 10^{-11},\label{Kpexp}\\
\B(K_L\to \pi^0\nu\bar\nu)_{\rm exp} &\leq 2.6\times 10^{-8}\qquad (90\%\, {\rm CL}).\label{KLexp}
\end{align}
The branching ratio of the charged mode is expected to be measured
with a precision of 10\%, relative to the SM prediction,  by the on-going 
NA62 experiment at CERN~\cite{Ruggiero:2017hjh}.
A search for the challenging neutral mode at the SM level 
is the ultimate goal of the KOTO experiment at JPARC~\cite{Beckford:2017qze}.

%%%%%%%%%%%%%%%%%%%%%%%%%%%%%%%%%%%%%%%%%%%%%%%%%%%%%%%%%%%
\boldmath
\section{The EFT approach to LFU violations based  on $U(2)_q\times U(2)_\ell$}
\label{sec:EFT}
\unboldmath
%%%%%%%%%%%%%%%%%%%%%%%%%%%%%%%%%%%%%%%%%%%%%%%%%%%%%%%%%%%

As already anticipated, the $B$-physics anomalies observed so far point toward NP 
coupled mainly to the third generation of SM fermions with some small (but non-negligible) mixing
with the light generations. In addition, all effects observed so far are well compatible with NP 
only involving left-handed currents. Left-handed four-fermion operators are also the most 
natural candidates to build a connection between anomalies in charged and neutral current 
semileptonic processes.
These observations have led to identify the EFT approach based on the $U(2)_q\times U(2)_\ell$
flavour symmetry as a a convenient framework (both successful and sufficiently general) 
to analyse $B$-physics anomalies and  discuss possible correlations with other low-energy 
observables~\cite{Greljo:2015mma,  Barbieri:2016las,Bordone:2017anc}.

The EFT is based on the assumption that the
first two generations of left-handed quarks and leptons transform as doublets of $U(2)_q\times U(2)_\ell$
while the third generation and the right-handed fermions are singlets
\begin{align}
Q&\equiv (q_L^1, q_L^2) \sim ({\bf 2},{\bf 1}), & q_L^3&\sim ({\bf 1},{\bf 1}),\\
L&\equiv (\ell_L^1,\ell_L^2)\sim ({\bf 1},{\bf 2}), & \ell_L^3&\sim({\bf 1},{\bf 1}).
\end{align}
Motivated by the observed pattern of the quark mass matrices, 
it is further  assumed that the leading breaking terms of this flavour 
symmetry are two spurion doublets, $V_q \sim ({\bf 2},{\bf 1})$ and $V_\ell \sim ({\bf 1},{\bf 2})$, that give rise to the mixing between 
the third generation and the other two~\cite{Barbieri:2011ci} (additional 
sub-leading breaking terms are needed to generate the masses of the light generations
and the corresponding mixing structures~\cite{Barbieri:2011ci}).

This symmetry and symmetry-breaking pattern implies $|V_{3i}| \approx |V_{i3}| \approx V_q^{(i)}$, up to model-dependent parameters of order one.
As a starting point, it is convenient to work in the down-quark mass basis, where the left-handed singlet and doublet fields read
\begin{align}
q_L^b &= \begin{pmatrix}V_{j3}^* u_L^j\\ b_L\end{pmatrix}, & Q_L^i &= \begin{pmatrix} V_{ji}^* u_L^j\\ d_L^i\end{pmatrix},\qquad (i = 1,2).
\end{align}
In this basis, one can set
\begin{equation}
V_q \propto \left(V_{td}^*,V_{ts}^*\right) \equiv \hat V_q,
\end{equation}
with the proportionality constant real and of order one. In the lepton sector, the size of the spurion $V_\ell$ is a free parameter, since it has no direct connection to the lepton 
Yukawa couplings.\footnote{It is worth stressing that in the lepton sector a different breaking pattern, i.e.~a leading breaking controlled by a triplet of $U(2)_\ell$, rather than a 
doublet, is also a viable option.}
Given that processes involving electrons are SM-like to a very high accuracy, we will assume $V_\ell = (0,\epsilon_\ell)$ with $|\epsilon_\ell | \ll 1$. 

The choice of the down-quark mass basis to identify singlets and doublets of the (quark) 
flavour symmetry is somehow arbitrary. In particular, the singlets do not need to be aligned with bottom 
quarks. On general grounds we expect 
\begin{align}
 q_{3L} &\equiv q_L^b + \theta_q e^{i\phi_q}\, {\hat V_q}^\dag\cdot Q_L~,
\label{q3}
\end{align}
were $\theta_q e^{i\phi_q}$ is the complex $O(1)$ parameter 
that  controls this possible mis-alignment:   $\theta_{q}\to0$ 
in case of alignment to the down-quark mass  basis, while 
$ \theta_q e^{i\phi_q} \to 1$ in the case of 
alignment to the up-quark mass basis.
Given the absence of deviations from the SM in CP-violating observables, 
it is natural to expect $\phi_q$ to be close to $0$ or $\pi$
($\theta_q $ is defined to be real and positive).
Similarly, in the lepton sector we define
\begin{equation}
\ell_{3L} \equiv \ell_L^3 + V_\ell^\dag \cdot L~.
\label{lepton_mixing}
\end{equation}

We shall describe NP effects through an EFT based on the following hypotheses:
\begin{enumerate}
\item the field content  below the NP scale $\Lambda > (G_F)^{-1/2}$ is the SM one;
\item the Lagrangian is invariant under the flavour symmetry $U(2)_q\times U(2)_\ell$, apart from the breaking induced by the spurions $V_q$ and $V_\ell$;
\item NP is directly coupled only to left-handed quark and lepton singlets in flavour space
(i.e.~only operators containing only $q_{3L}$ or $\ell_{3L}$  fields are affected by tree-level matching conditions at the NP scale $\Lambda$). 
\end{enumerate}

Given these assumptions,  we can identify only two independent operators of dimension six affected by NP and 
contributing to semileptonic decays at the tree level, namely the electroweak singlet and triplet current-current interactions,
\begin{equation}\label{LNP}
\L_{\rm eff} = -\frac{1}{\Lambda^2}(\bar{q}_{3L} \gamma_\mu \sigma^a q_{3L})(\bar{\ell}_{3L} \gamma^\mu \sigma^a \ell_{3L}) - \frac{c_{13}}{\Lambda^2}(\bar{ q}_{3L} \gamma_\mu  q_{3L})(\bar{ \ell}_{3L} \gamma^\mu \ell_{3L})~.
\end{equation}
The normalisation of the triplet operator in (\ref{LNP}) has been chosen in order to generate a constructive interference with the SM in 
charged-current amplitudes, as suggested by $b\to c \tau \bar\nu_\tau$ data. The overall-scale of this operator defines 
the NP scale $\Lambda$, while $c_{13}$ denotes the ratio between the singlet and triplet Wilson coefficients.

%%%%%%%%%%%%%%%%%%%%%%%%%%%%%%%%%%%%%%%%%%%%%%%%%%%%%%%%%%%
\section{Physical observables}\label{sec:obs}
%%%%%%%%%%%%%%%%%%%%%%%%%%%%%%%%%%%%%%%%%%%%%%%%%%%%%%%%%%%

%%%%%%%%%%%%%%%%%%%%%%%%%%%%%%%%%%%%%%%%%%%%%%%%%%%%%%%%%%%
\boldmath
\subsection{The $R_{D^{(*)}}$ anomaly}\label{sec:RD}
\unboldmath
%%%%%%%%%%%%%%%%%%%%%%%%%%%%%%%%%%%%%%%%%%%%%%%%%%%%%%%%%%%

The averages of the  $\tau/\ell$ universality  ratios ($\ell={\mu,e}$)
in $b\to c$  transitions measured by BaBar~\cite{Lees:2013uzd}, Belle~\cite{Hirose:2016wfn}
and LHCb~\cite{Aaij:2015yra}, are  
\begin{align}
R_{D^*} &\equiv \frac{\B(B\to D^* \tau \bar\nu_\tau)_{\rm exp}/\B(B\to D^* \tau \bar\nu_\tau)_{\rm SM}}{\B(B\to D^* \ell \bar\nu_\ell)_{\rm exp}/\B(B\to D^* \ell \bar\nu_\ell)_{\rm SM}} = 1.23\pm 0.07\,,\\
R_{D} &\equiv \frac{\B(B\to D \tau \bar\nu_\tau)_{\rm exp}/\B(B\to D \tau \bar\nu_\tau)_{\rm SM}}{\B(B\to D \ell \bar\nu_\ell)_{\rm exp}/\B(B\to D \ell \bar\nu_\ell)_{\rm SM}} = 1.35\pm 0.16\,.
\end{align}
These two results can be combined into a single observable that parametrises the violation of LFU in charged currents
(assuming a purely left-handed structure): 
\begin{equation}
R_{D^{(*)}} = 1.24\pm0.07.
\end{equation}

Only the triplet operator in \eqref{LNP} contributes to $b\to c\tau\bar\nu_{\tau}$ decays via the following effective interaction 
\begin{equation}
\L_{b\to c\tau\bar\nu_{\tau}}^{\rm NP} =
-\frac{2}{\Lambda^2}\Big[V_{cb} +  \theta_q e^{-i\phi_q} (V_{cs}V_{ts}^* + V_{cd}V_{td}^*) \Big]
(\bar{c}_L\gamma^{\mu}b_{L})(\bar{\tau}_L\gamma_{\mu}\nu_{\tau})\,.
\end{equation}
The branching ratio for the processes $B\to D^{(*)}\tau\bar\nu$  is then modified as follows by the triplet operator (using CKM unitarity and setting $V_{tb} = 1$)
\begin{align}
\B(B\to D^{(*)}\tau\bar\nu) = \B(B\to D^{(*)}\tau\bar\nu)_{\rm SM}\left| 1 + R_0 \left( 1 - \theta_q e^{-i\phi_q} \right) \right|^2
\end{align}
where we have defined 
\begin{equation}
\displaystyle{R_0 = \frac{1}{\Lambda^2}\frac{1}{\sqrt{2}G_F}}.
\end{equation} 
In the limit where we neglect sub-leading terms suppressed by the small leptonic spurion, 
NP does not affect  $\B(B\to D^{(*)}\ell\bar\nu)$  for the light leptons.
This allows us to fix the overall scale of NP via the relation
\begin{equation}
\left[R_{D^{(*)}}^{\tau/\mu}-1\right] \approx 2R_0(1-\theta_q \cos\phi_q) = 0.24\pm 0.07\,.\label{RD}
\end{equation}
The reference effective scale of  NP, obtained for 
$\theta_q \to 0$, is $\Lambda_0  \approx 700\,{\rm GeV}$. 
Notice that higher scales of NP can be obtained if $\theta_q =O(1)$ and 
$\cos\phi_q < 0$,
obtaining in this way a better compatibility with constraints from direct searches~\cite{Faroughy:2016osc}
and electroweak precision tests~\cite{Feruglio:2017rjo,Feruglio:2016gvd}. 
On the other hand, the NP contribution to $R_{D^{(*)}}$ vanishes in the case 
of alignment of the flavour symmetry to up-type quarks ($\theta_q\to 1, \phi_q\to 0$).

%%%%%%%%%%%%%%%%%%%%%%%%%%%%%%%%%%%%%%%%%%%%%%%%%%%%%%%%%%%
\boldmath
\subsection{LFU violating contributions to $K\to \pi\nu\bar\nu$}\label{sec:kpnnLFU}
\unboldmath
%%%%%%%%%%%%%%%%%%%%%%%%%%%%%%%%%%%%%%%%%%%%%%%%%%%%%%%%%%%

\begin{figure}
\centering%
\includegraphics[width=0.49\textwidth]{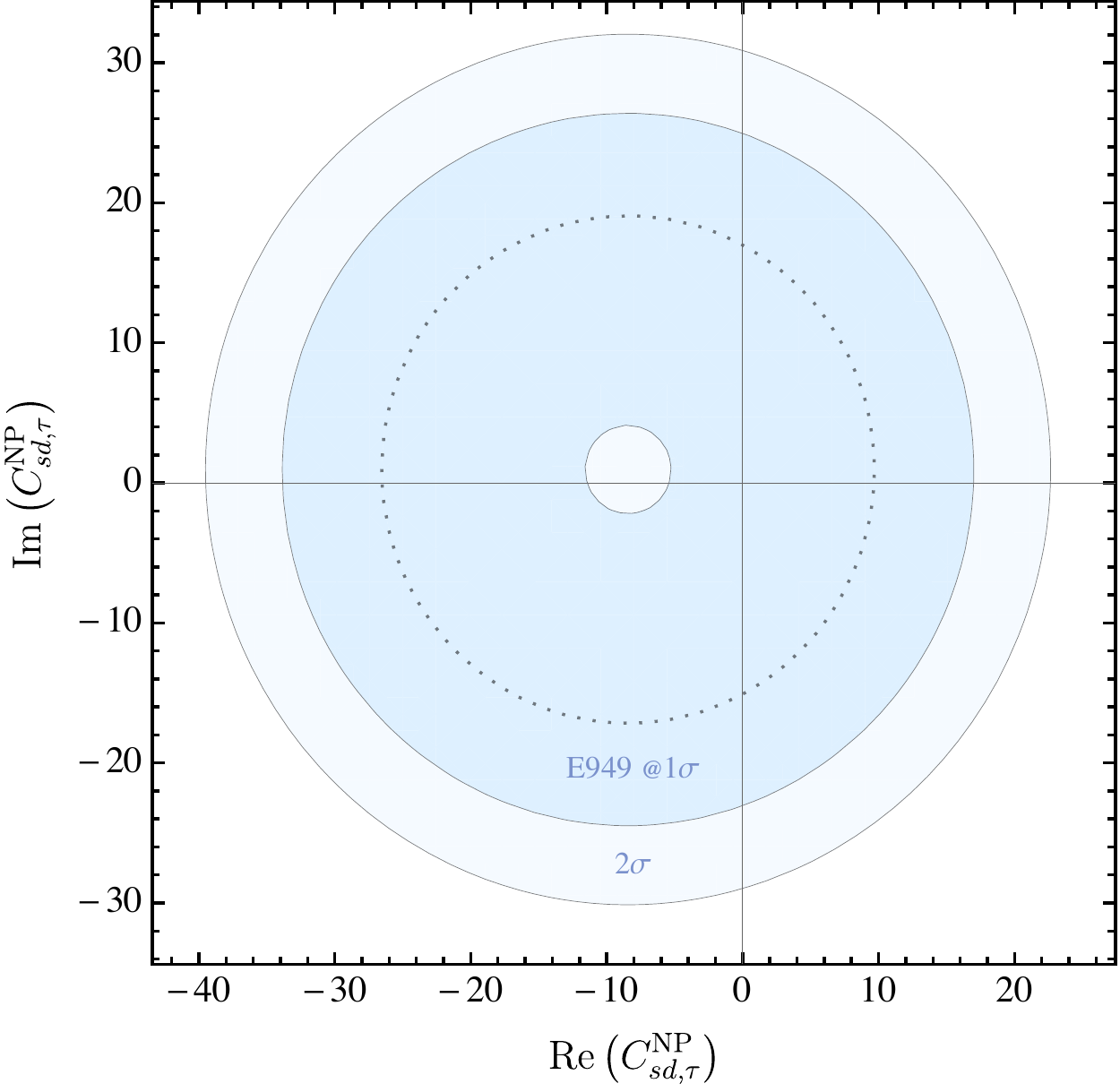}\hfill%
\raisebox{0.06em}{\includegraphics[width=0.485\textwidth]{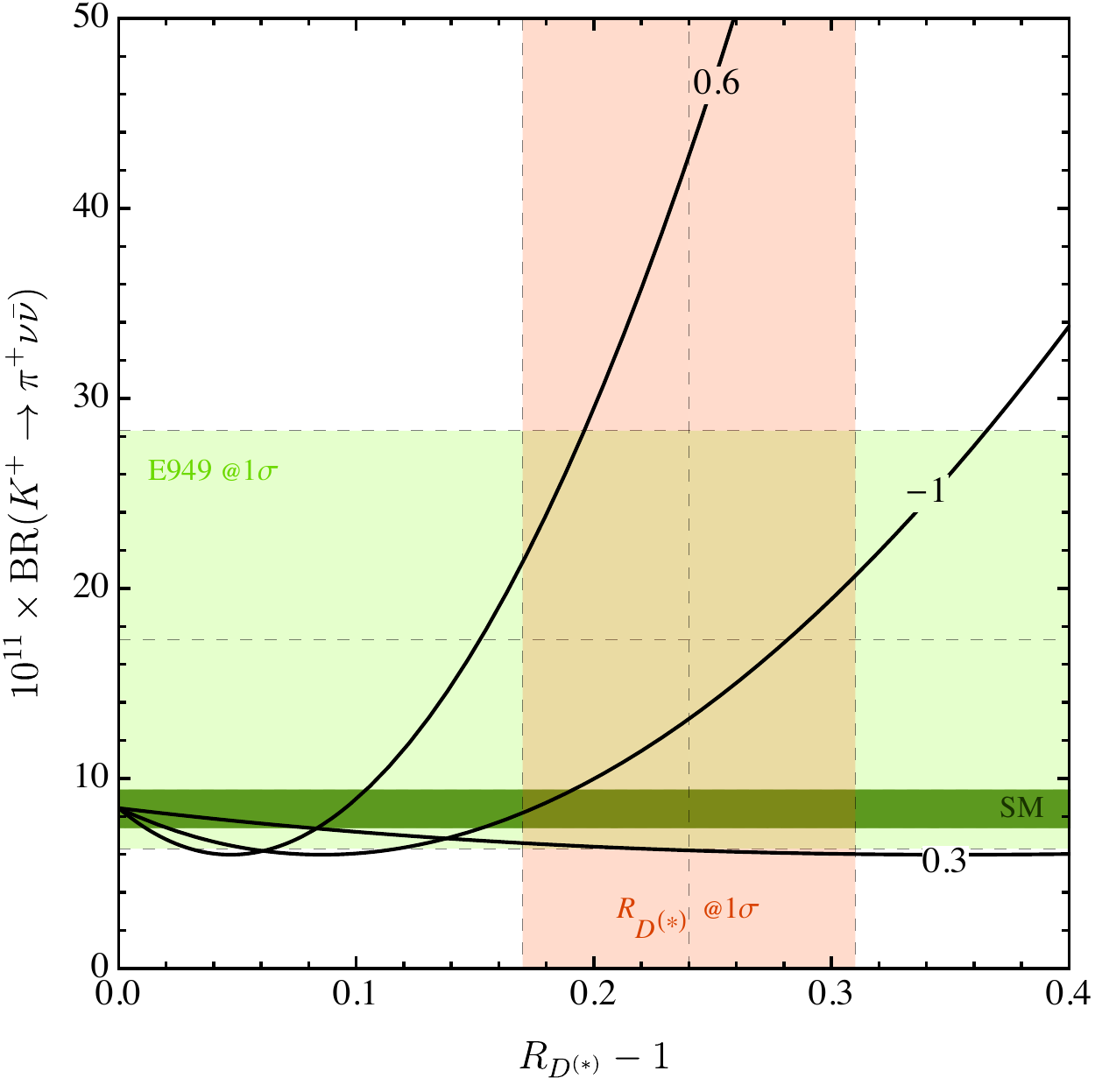}}
\caption{Left: allowed range for the real and imaginary parts of the NP Wilson coefficient $C_{sd,\tau}^{\rm NP}$. Right: correlation between $\B(K^+\to\pi^+\nu\bar\nu)$ and $R_{D^{(*)}}$ for different values of the parameter $\theta_q$ (with $\phi_q = c_{13} = 0$); the coloured regions are the experimental measurements at $1\,\sigma$, the dark green band is the SM prediction.\label{fig:1}}
\end{figure}

The operators \eqref{LNP} contribute to $s\to d\nu\bar\nu$ transitions through the term proportional to the $V_q$ spurion in \eqref{q3},
\begin{equation}
\L_{s\to d\nu\bar\nu}^{\rm NP} = \frac{1-c_{13}}{\Lambda^2}\theta_q^2\, V_{ts}^*V_{td}(\bar s_L\gamma_\mu d_L)(\bar\nu_\tau\gamma_\mu\nu_\tau).
\end{equation}
Neglecting, in first approximation, the NP contribution to $s\to d\nu_\ell\bar\nu_\ell$ ($\ell=e,\mu$) amplitudes, 
we can write
\begin{eqnarray}
\B(K^+\to \pi^+\nu\bar\nu) &=& 2\B(K^+\to\pi^+\nu_e\bar\nu_e)_{\rm SM} + \B(K^+\to\pi^+\nu_\tau\bar\nu_\tau)_{\rm SM}\left| 1 + \frac{R_0\, \theta_q^2(1-c_{13})}{(\alpha/\pi) C_{sd,\tau}^{\rm SM,eff}}  
\right|^2~,  \nonumber \\
\B(K_L \to \pi^0 \nu\bar\nu) &=& 2\B(K_L \to\pi^0 \nu_e\bar\nu_e)_{\rm SM} + \B(K_L \to\pi^0 \nu_\tau\bar\nu_\tau)_{\rm SM}\left| 1 - \frac{R_0\, \theta_q^2(1-c_{13})}{(\alpha/\pi) (X_t/ s_w^2) }  
\right|^2~,  \nonumber \\ 
\end{eqnarray}
where $C_{sd,\tau}^{\rm SM,eff} \approx  -8.5 \times e^{0.11 i}$ includes also the long-distance contributions of \eqref{Kpvv}.

The current allowed range from the experimental result \eqref{Kpexp} for the real and imaginary parts of the Wilson coefficient $C_{sd,\tau}^{\rm NP}$ in a generic NP model is shown in Figure~\ref{fig:1} (left).
In our case this translates into the constraint
\begin{equation}
|R_0\, \theta_q^2(1-c_{13})|  \lesssim 0.1~.
\label{eq:constr_theta}
\end{equation}
As expected, the constraint vanishes in the limit $c_{13} \to 1$, where triplet and singlet  NP contributions 
to $s\to d\nu\bar\nu$ amplitudes cancel each other.
However, it must be stressed that there is no symmetry reason to expect $c_{13} = 1$. Even if   $c_{13} = 1$ holds 
as tree-level matching condition in the EFT (such as e.g.~in the lepto-quark models of Ref.~\cite{Barbieri:2015yvd,Crivellin:2017zlb}), 
one expects $c_{13} \not= 1$ beyond the tree level~\cite{Barbieri:2015yvd}. For $c_{13} \not= 1$ the result in 
\eqref{eq:constr_theta} implies a severe constraint on the maximal value of $\theta_q$,
assuming \eqref{RD} is satisfied. For $|c_{13} - 1| \ll 1$ one finds $|\theta_q  |  \lesssim 1/ |c_{13} - 1|$.

\begin{figure}
\centering%
\includegraphics[width=0.485\textwidth]{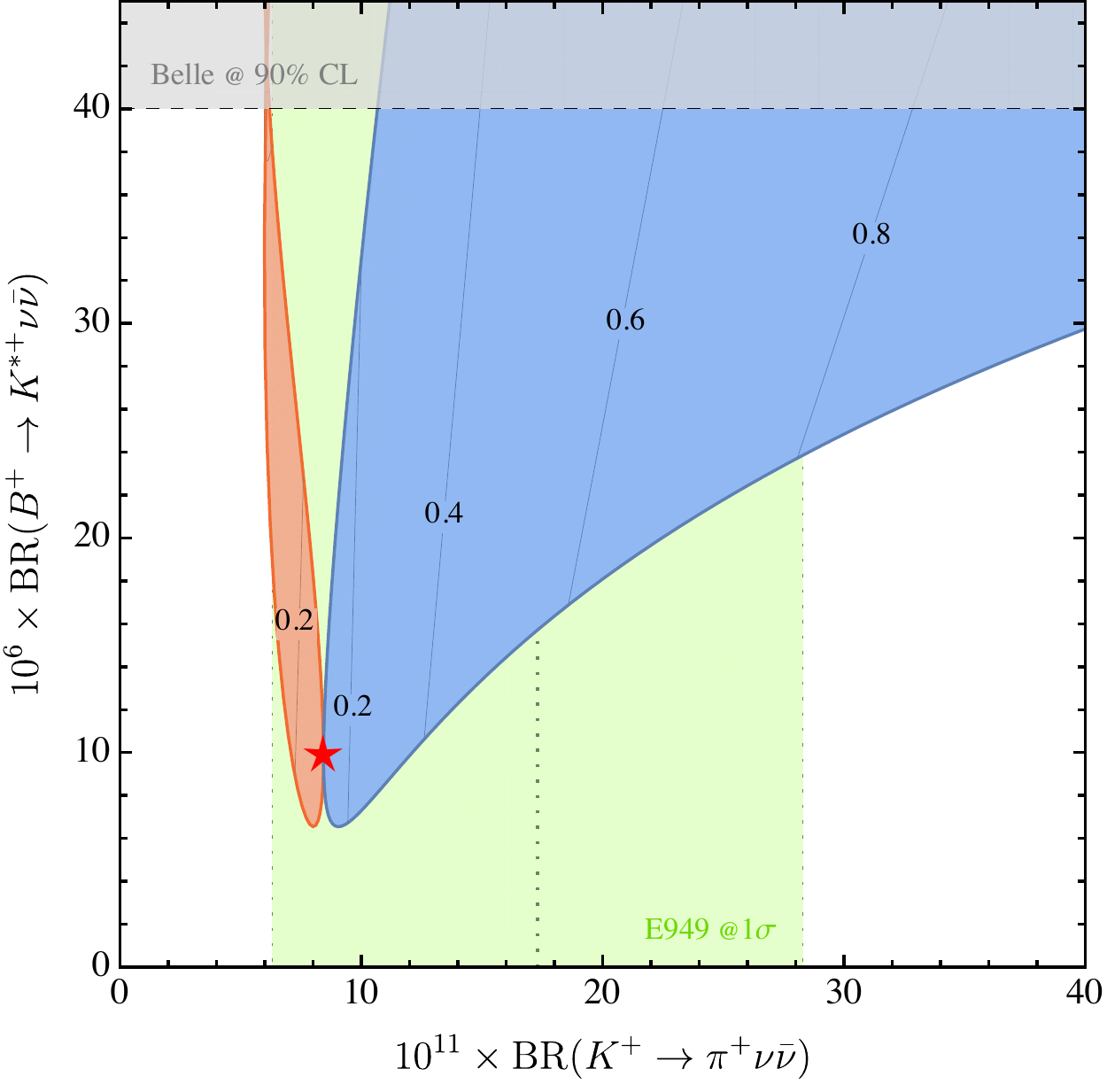}\hfill%
\raisebox{0.05em}{\includegraphics[width=0.47\textwidth]{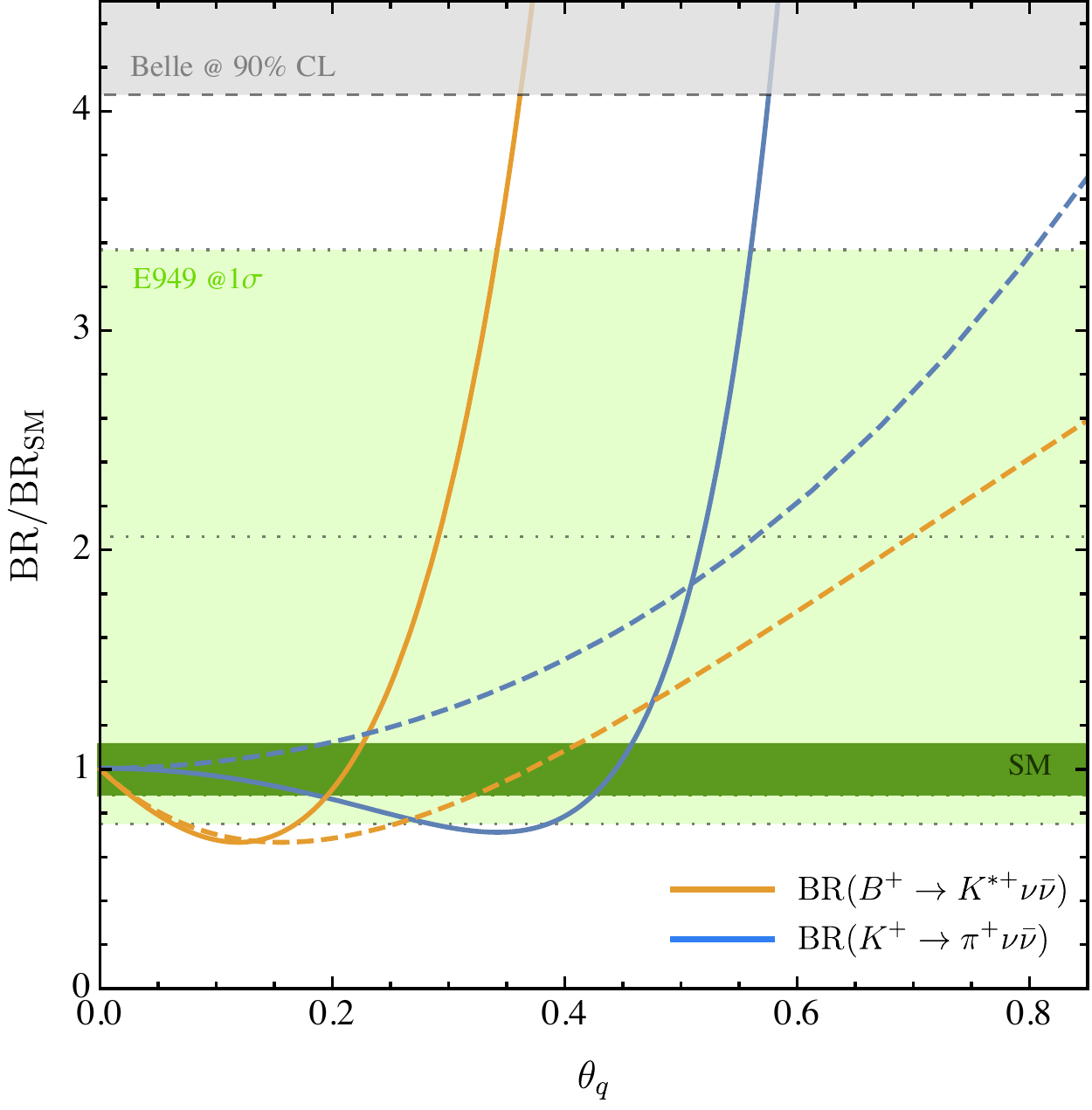}}
\caption{Left: correlation of $\B(K^+\to \pi^+\nu\bar\nu)$ with $\B(B^+\to K^{*+}\nu\bar\nu)$, having imposed $R_{D^{(*)}} = 1.25$. The red (blue) colored region is for $c_{13} = 0$ ($c_{13} = 2$). We also show isolines of $\theta_q$, and the red star is the SM point. Right: branching ratios for $K^+\to\pi^+\nu\bar\nu$ and $B^+\to K^{*+}\nu\bar\nu$, normalised to the SM values, as functions of $\theta_q$. The solid (dashed) lines correspond to $c_{13} = 0$, $\phi_q = 0$ ($c_{13} = 2$, $\phi_q = \pi$).\label{fig:2}}
\end{figure}

Expressing $R_0$ in terms of the measured value of $R_{D^{(*)}}$ (and the unknown parameters $\theta_q$ and $\phi_q$)
we can rewrite the previous expression as a relation between $R_{D^{(*)}}$ and $\B(K\to \pi\nu\bar\nu)$ as follows
\begin{align}
\B(K^+\to \pi^+\nu\bar\nu) &\approx \B(K^+\to \pi^+\nu\bar\nu)_{\rm SM} \Big[ 1 
 - 14 \, [R_{D^{(*)}}-1]\theta^2_q f_q + 165\, [R_{D^{(*)}}-1]^2\theta_q^4 f_q^2\Big]\,,
\end{align}
where
$f_q \equiv   (1-c_{13}) / (1-\theta_q\cos\phi_q)$,
 and where we neglected higher orders in $R_0$ from \eqref{RD}.
This correlation is shown in Figure~\ref{fig:1} (right), for different values of the free parameters. As can be seen, for $\theta_q=O(1)$ the solution of the $R_{D^{(*)}}$ anomaly can imply sizeable deviations in $\BK$  compared to the SM.
The dependence of $\BK$ on the parameter $\theta_q$, with $\RD$ fixed as in \eqref{RD}, is shown by the blue lines in Figure~\ref{fig:2} (right) for the two values of the phase $\phi_u = 0$ and $\pi$,   and for different values of the singlet contribution $c_{13}$. Notice that for $c_{13}>1$ the branching ratio is always enhanced with respect to the SM prediction.

The neutral mode $K_L\to \pi^0\nu\bar\nu$ is purely CP-violating and constrains only the imaginary part of the amplitude. The present bound on the NP Wilson coefficient from \eqref{KLexp} is roughly 10 times weaker than the one from the $K^+$ mode.

%%%%%%%%%%%%%%%%%%%%%%%%%%%%%%%%%%%%%%%%%%%%%%%%%%%%%%%%%%%
\boldmath
\subsection{Correlations between $B\to K^{(*)}\nu\bar\nu$ and $K\to \pi\nu\bar\nu$}\label{sec:bknn}
\unboldmath
%%%%%%%%%%%%%%%%%%%%%%%%%%%%%%%%%%%%%%%%%%%%%%%%%%%%%%%%%%%

Also $b\to s\nu\bar\nu$ transitions are described by the Lagrangian \eqref{Leff}, with $C^{\rm SM}_{bs} = -X_t/s_w^2\approx -6.4$. Notice that the charm contribution is not relevant in this case. Both charged and neutral $B\to K^{(*)}\nu\bar\nu$ decays set bounds on the New Physics Wilson coefficient, with the stronger constraints coming from the $B^+$ modes.
In the SM, the branching ratios are~\cite{Buras:2014fpa}
\begin{align}
\B(B^+\to K^+\nu\bar\nu)_{\rm SM} &= 3.94\times 10^{-6}\left|\frac{V_{ts}V_{tb}^*}{0.04}\right|^2,\label{BKp}\\
\B(B^+\to K^{+*}\nu\bar\nu)_{\rm SM} &= 9.82\times 10^{-6}\left|\frac{V_{ts}V_{tb}^*}{0.04}\right|^2,\label{BK0}
\end{align}
to be compared with the experimental bounds~\cite{Olive:2016xmw}
\begin{align}
\B(B^+\to K^+\nu\bar\nu) &\leq 1.7\times 10^{-5}\quad \text{@ 90\% CL}~,\\
\B(B^+\to K^{+*}\nu\bar\nu) &\leq 4.0\times 10^{-5}\quad \text{@ 90\% CL}~.
\end{align}
In the presence of the operators of \eqref{LNP} the branching ratios are modified as follows
\begin{equation}
\B(B\to K^{(*)}\nu\bar\nu) = \B(B\to K^{(*)}\nu\bar\nu)_{\rm SM}\left[\frac{2}{3} + \frac{1}{3}\left| 1 -  \frac{\pi s_w^2 R_0}{\alpha X_t}\theta_q e^{i\phi_q}(1-c_{13})\right|^2\right]~.
\end{equation}
As for $K\to\pi\nu\bar\nu$, we can obtain  a direct connection to the charged-current anomaly
expressing $R_0$ in terms of  $R_{D^{(*)}}$:
\begin{align}
\B(B\to K^{(*)}\nu\bar\nu) &\approx \B(B\to K^{(*)}\nu\bar\nu)_{\rm SM}\Big(1 - 21 [R_{D^{(*)}}-1] \theta_q \cos\phi_q f_q + 320 [R_{D^{(*)}}-1]^2 \theta_q^2 f_q^2\Big).
\end{align}
Figure~\ref{fig:2} (right) shows the dependence of $\B(B^+\to K^{*+}\nu\bar\nu)$ on $\theta_q$, with $\RD$ fixed to the central value of the experimental measurement, and for two different values of $\phi_q$ and $c_{13}$.

The deviations from the SM expectations in the two FCNC neutrino modes are closely correlated, as described by the following relation
\begin{align}
\frac{ \Delta\B(K^+ \to \pi^+\nu\bar\nu) }{ \Delta  \B(B \to K^{(*)} \nu\bar\nu) } \approx \frac{2}{3} \times
\frac{\theta_q}{ \cos\phi_q} \times \frac{1 - 12\,  [R_{D^{(*)}}-1] \theta^2_q f_q}{1 - 15 [R_{D^{(*)}}-1]\, \frac{\theta_q f_q}{ \cos\phi_q} }\,,
\end{align}
where $\Delta\B = \frac{\B - \B_{\rm SM}}{\B_{\rm SM}}$, and as illustrated 
in Figure~\ref{fig:2}. Notice that for small $\theta_q$ this correlation does not depend on the measured value of $\RD$.
The constraints from $B\to K^{(*)}\nu\bar\nu$ can severely limit the deviations in $K\to \pi\nu\bar\nu$. If $c_{13} <1$, the NP contributions interfere constructively with the SM amplitude in the first branching ratio, and destructively in the second one. As a consequence, in this case $\BK$ is always suppressed, with deviations of up to $-30\%$ with respect to the SM value. The opposite is true when $c_{13}>1$. Also, the constraints are more stringent when $\cos\phi_q$ is positive, since in this case the effective scale of new physics is lower. For negative $\cos\phi_q$ and $c_{13}>1$, in particular, the constraint from $B\to K^{(*)}\nu\bar\nu$ becomes irrelevant, and large deviations can be expected in $\BK$ (within the limits of \eqref{Kpexp}).

%%%%%%%%%%%%%%%%%%%%%%%%%%%%%%%%%%%%%%%%%%%%%%%%%%%%%%%%%%%
\subsection{Constraints and connections to other observables}\label{sec:constraints}
%%%%%%%%%%%%%%%%%%%%%%%%%%%%%%%%%%%%%%%%%%%%%%%%%%%%%%%%%%%

\paragraph{\boldmath$b\to s \ell^+\ell^-$.\unboldmath}

FCNC processes that involve the light generations of leptons are suppressed by the spurion $V_\ell$ in our framework.
While LFU violation in these modes is a general prediction following from \eqref{lepton_mixing}, the exact size of these effects depends on the unknown parameter $\epsilon_\ell$ 
(and, more generally, by the assumption on the breaking of the $U(2)_\ell$ lepton flavour symmetry). 
The NP contributions to the Wilson coefficients $C_9$ and $C_{10}$ of the semileptonic $b\to s\mu^+\mu^-$ Lagrangian
\begin{equation}
\L_{\rm eff}^{b\to s\mu\mu} = \frac{4G_F}{\sqrt{2}}\frac{\alpha}{4\pi}V_{tb}^*V_{ts}\left[C_{9,\mu} (\bar b_L\gamma_\mu s_L)(\bar\mu\gamma^\mu \mu) + C_{10,\mu}(\bar b_L\gamma_\mu s_L)(\bar\mu \gamma^\mu\gamma_5 \mu)\right]
\end{equation}
read
\begin{equation}
C_{9,\mu}^{\rm NP} = -C_{10,\mu}^{\rm NP} = -\frac{\pi}{\alpha}R_0 \theta_q e^{i\phi_q} (1+c_{13})|\epsilon_\ell|^2~.\label{C9C10}
\end{equation}
Global fits of these Wilson coefficients, performed after the recent measurement of the LFU ratio $R_{K^{*}}$~\cite{Aaij:2017vbb},
in the case of NP coupled to left-handed currents only, yields 
$C_{9,\mu}^{\rm NP} = -C_{10,\mu}^{\rm NP} = -0.64\pm 0.18$~\cite{Altmannshofer:2017yso,Capdevila:2017bsm,Geng:2017svp,Ciuchini:2017mik,DAmico:2017mtc}.
From this result, fixing the overall scale of NP from \eqref{RD}, it follows that $|\epsilon_\ell|\,(1+c_{13}) \approx 0.1$ up to an $O(1)$ 
 factor depending on $\theta_q$ and $\phi_q$. Since the sign of $C_9^{\rm NP}$ must be negative to fit the $b\to s\ell\bar\ell$ anomalies, 
it follows that $(1-c_{13}) \cos\phi_q > 0$.

\paragraph{\boldmath$b\to s \tau^+\tau^-$.\unboldmath} 

FCNC decays of $B$ mesons with a $\tau^+\tau^-$ pair in the final state arise at leading order in the breaking of $U(2)_q\times U(2)_\ell$. This implies that 
these processes can be directly related both to the $\RD$ anomalies, 
and to the two neutrino modes discussed above. The current experimental limits on $\B(B\to K\tau^+\tau^-)$ are four orders of magnitude larger than 
the corresponding SM prediction (which lies in the $10^{-7}$ range). 
The Belle II experiment is expected to improve these limits by at least one order of magnitude, reaching the $10^{-4}$ level~\cite{Zupanc:2017CERN}.
While the value predicted in the SM would still be out of reach, this sensitivity could be interesting in the NP framework introduced above. 
The relevant NP Wilson coefficient are $C_{9,\tau}^{\rm NP} =  -C_{10,\tau}^{\rm NP} = C_{9,\mu}^{\rm NP}/|\epsilon_\ell|^2$, and
the branching ratio depends quadratically on $R_0 (1+c_{13})$ in the limit where the NP contribution is large. Setting 
$\RD$ to the central value in \eqref{RD}, and imposing the constraints on $\theta_q$ from $\B\to K^{(*)}\nu\bar\nu$ and $K\to \pi\nu\bar\nu$, 
one gets an enhancement of a factor $10^{2}\div 10^{3}$ in  $\B(B\to K\tau^+\tau^-)$ 
 if $\left| \frac{1-c_{13}}{1+c_{13}}\right| \lesssim 20\%$ (which is a rather natural 
choice of parameters).
Finally, it is interesting to note that the observation of $b\to s\tau^+\tau^-$ transitions, together with 
$s\to d\nu\bar\nu$ and $b\to s\nu\bar\nu$, would allow to fix the three dimensionless parameters $c_{13}$, $\theta_q$, and $\phi_q$ 
entering the Lagrangian \eqref{LNP}, thus completely determining the leading free parameters of the EFT.

\paragraph{\boldmath $\tau\to K \nu$.\unboldmath}\label{tau}

The $s\to u$ analogue of $B\to D^{(*)}\tau\bar\nu$ is the tau decay $\tau \to K \nu$, which is generated at tree-level in the SM, and gets a contribution from the charged-current interaction \eqref{LNP}. The total branching ratio in presence of NP can be written as
\begin{equation}
\B(\tau\to K\nu) = \B(\tau\to K\nu)_{\rm SM}\left| 1 - R_0 \frac{V_{ub}V_{ts}}{\lambda} \theta_q (e^{i\phi}-\theta_q)\right|^2.
\end{equation}
The measured value is $\B(\tau\to K\nu)_{\rm exp} = (6.9\pm 0.1)\times 10^{-3}$, which has to be compared with the SM prediction $\B(\tau\to K\nu)_{\rm SM} = (7.1\pm 0.1)\times 10^{-3}$. This translates into a loose bound on the scale of NP [$R_0\theta_q(1-\theta_q) \lesssim 20$ for  $\phi_q = 0$].

\paragraph{Loop effects.}

The running from the scale $\Lambda$ to the electroweak scale, starting from the NP semileptonic Lagrangian \eqref{LNP},
does generate non-vanishing contributions to four-quark and four-lepton operators. 
The contributions to $K-\bar K$ and $B_s-\bar B_s$ mixing, as well as to flavour-changing $Zq\bar q$ interactions, 
are suppressed at least by the $\tau$ mass, and turn out to be several orders of magnitude below present experimental constraints.

It is on the other hand known that running effects due to quark loops, leading to purely leptonic operators~\cite{Feruglio:2017rjo,Feruglio:2016gvd},
are potentially more problematic because of precise constraints from leptonic $\tau$ decays.
In concrete models, additional UV contributions to the same effective operators will arise from the matching at the scale $\Lambda$. These contributions can help satisfying the $\tau$ decay constraints, but can also constitute a problem for meson mixing. On general grounds, satisfying all the constraints in a concrete UV completion that incorporates both the $b\to c\tau\bar\nu$ and $b\to s\ell^+\ell^-$ anomalies is not straightforward. 
However, as shown in~\cite{Bordone:2017anc}, this result can be achieved with a moderate tuning of parameters. 
Given the model-dependence of the radiative constraints, we do not take them into account in the present analysis whose main focus are semileptonic decays.

%%%%%%%%%%%%%%%%%%%%%%%%%%%%%%%%%%%%%%%%%%%%%%%%%%%%%%%%%%%
\section{Conclusions}\label{sec:conclusions}
%%%%%%%%%%%%%%%%%%%%%%%%%%%%%%%%%%%%%%%%%%%%%%%%%%%%%%%%%%%

Recent $B$-physics data  hints toward violations of  Lepton Flavour Universality  in charged-  and neutral-current 
semileptonic processes.  The most natural explanation of these phenomena, if both will be confirmed as evidences of physics  beyond the SM,
is the hypothesis of a new interaction in the TeV range that couples mainly to third-generation fermions. 
If a CKM-like relation connects NP effects in $B$ and $K$ physics, it is natural to expect sizeable deviations from the SM in $K\to \pi\nu\bar\nu$ decays, 
which are the only $s\to d$ transitions that involve third-generation leptons in the final state.

To quantify possible NP effects in  $K\to \pi\nu\bar\nu$ decays in sufficiently general terms, being motivated by present $B$-physics anomalies, we have 
considered an EFT based on the hypothesis of a $U(2)_q\times U(2)_\ell$  flavour symmetry acting on the light generations of left-handed fermions,
broken in the quark sector by the small CKM-like spurion $V_q$ connecting third and light generations (and similarly broken by a small spurion $V_\ell$
in the lepton sector).
We further assumed that NP is 
 coupled only    to the left-handed third generation flavour-singlets ($q_{3L}$ and $\ell_{3L}$). 
Because of the freedom in the choice of the flavour basis, 
the spurions $V_{q,\ell}$ can enter the definition of the flavour singlets with an arbitrary mixing parameter of order one. The latter 
control the communication of NP effects from processes with third-generation fermions only, to processes with light generations. 
This set-up is not the most general one compatible with the $U(2)_q\times U(2)_\ell$  flavour symmetry, but it covers a wide class 
of the most motivated explicit models so far proposed to address $B$-physics anomalies.

In this framework, we focused our attention to semileptonic transitions involving only $\tau$ leptons and $\tau$ neutrinos.
These processes are completely determined by four real parameters: the overall scale of the new interactions $\Lambda$, the two model-dependent real (mixing) 
parameters $\theta_q$ and $\phi_q$  defining the (quark) flavour basis, and the relative strength of the electroweak-triplet and -singlet NP interactions $c_{13}$.
The measurement of the LFU ratios $R_{D^{(*)}}$ can be used to fix the NP scale $\Lambda$ in terms of $\theta_q$ and $\phi_q$. This allows in turn to 
study the neutrino FCNC transitions $K\to\pi\nu\bar\nu$ and $B\to K^{(*)}\nu\bar\nu$, as well as $B\to K^{(*)}\tau^+\tau^-$,
as functions of the three remaining parameters naturally expected to be of $O(1)$.

We have shown that, for natural values of the free parameters, sizeable and closely correlated deviations from the SM of both neutrino modes are expected.
The electroweak triplet operator alone necessarily causes a suppression of $\BK$, due to the interference of NP with the SM amplitude which is always destructive. 
This suppression could be as large as $30\%$, relative the SM value. 
If, on the other hand, also an electroweak singlet interaction is present, arbitrary modifications of $\BK$ are possible.
The strongest constraint on the allowed size of these deviations comes from the present bounds on $\BB$ which, however, 
do not exclude $O(1)$ enhancements in $\BK$, as illustrated in Figure~\ref{fig:2}.

Order of magnitude enhancements of $b\to s\tau^+\tau^-$ compared to the SM 
are possible in this class of NP models. However, these transitions are very challenging from the experimental point of view.  In principle, 
the combined measurement of $\RD$, $\BK$, $\BB$, and $\B(B\to K\tau^+\tau^-)$ would allow to completely determine the leading 
parameters of the EFT.  The correlation with other observables is less straightforward: violations of $\mu/e$ universality in $b\to s\ell\bar\ell$ transitions are a natural prediction of this framework; however, their size and the correlation with 
NP effects in the neutrino modes are controlled by additional free parameters.
 
Summarising,  $K\to\pi\nu\bar\nu$ decays could be significantly affected by the non-standard LFU-violating interactions  
hinted by present $B$-physics data. The forthcoming measurement of $\B(K^+\to \pi^+\nu\bar\nu)$
by the NA62 experiment at CERN will  provide an important insight on this class of NP models. 
The general expectation is a sizeable deviation from the SM, that, however, could result also into a significant suppression. 
Should a deviation from the SM prediction be observed in this channel, its correlation with NP effects in $\BB$  and, possibly, 
$\B(B\to K\tau^+\tau^-)$, would allow to reveal the flavour structure of this new interaction.

%%%%%%%%%%%%%%%%%%%%%%%%%%%%%%%%%%%%%%%%%%%%%%%%%%%%%%%%%%%
\subsubsection*{Acknowledgements}
%%%%%%%%%%%%%%%%%%%%%%%%%%%%%%%%%%%%%%%%%%%%%%%%%%%%%%%%%%%

We thank Admir Greljo and David Marzocca for many useful discussions. This research was supported in part by the Swiss National Science Foundation (SNF) under contract 200021-159720.

%%%%%%%%%%%%%%%%%%%%%%%%%%%%%%%%%%%%%%%%%%%%%%%%%%%%%%%%%%%
%%%%%%%%%%%%%%%%%%%%%%%%%%%%%%%%%%%%%%%%%%%%%%%%%%%%%%%%%%%

\bibliographystyle{my.bst}

{\footnotesize
\bibliography{arxiv_v1}

\providecommand{\href}[2]{#2}\begingroup\raggedright\begin{thebibliography}{10}

\bibitem{Aaij:2017vbb}
{\bf LHCb}, R.~Aaij {\em et al.,}
\href{http://arxiv.org/abs/1705.05802}{{\tt arXiv:1705.05802 [hep-ex]}}.
%%CITATION = ARXIV:1705.05802;%%.

\bibitem{Lees:2013uzd}
{\bf BaBar}, J.~P. Lees {\em et al.,}
  \href{http://dx.doi.org/10.1103/PhysRevD.88.072012}{{\em Phys. Rev.} {\bf
  D88} (2013) no.~7, 072012},
\href{http://arxiv.org/abs/1303.0571}{{\tt arXiv:1303.0571 [hep-ex]}}.
%%CITATION = ARXIV:1303.0571;%%.

\bibitem{Hirose:2016wfn}
{\bf Belle}, S.~Hirose {\em et al.,}
\href{http://arxiv.org/abs/1612.00529}{{\tt arXiv:1612.00529 [hep-ex]}}.
%%CITATION = ARXIV:1612.00529;%%.

\bibitem{Aaij:2015yra}
{\bf LHCb}, R.~Aaij {\em et al.,}
  \href{http://dx.doi.org/10.1103/PhysRevLett.115.159901,
  10.1103/PhysRevLett.115.111803}{{\em Phys. Rev. Lett.} {\bf 115} (2015)
  no.~11, 111803}, \href{http://arxiv.org/abs/1506.08614}{{\tt arXiv:1506.08614
  [hep-ex]}}.
[Addendum: Phys. Rev. Lett.115,no.15,159901(2015)].
%%CITATION = ARXIV:1506.08614;%%.

\bibitem{Aaij:2014ora}
{\bf LHCb}, R.~Aaij {\em et al.,}
  \href{http://dx.doi.org/10.1103/PhysRevLett.113.151601}{{\em Phys. Rev.
  Lett.} {\bf 113} (2014)  151601},
\href{http://arxiv.org/abs/1406.6482}{{\tt arXiv:1406.6482 [hep-ex]}}.
%%CITATION = ARXIV:1406.6482;%%.

\bibitem{Altmannshofer:2017yso}
W.~Altmannshofer, P.~Stangl, and D.~M. Straub,
\href{http://arxiv.org/abs/1704.05435}{{\tt arXiv:1704.05435 [hep-ph]}}.
%%CITATION = ARXIV:1704.05435;%%.

\bibitem{Capdevila:2017bsm}
B.~Capdevila, A.~Crivellin, S.~Descotes-Genon, J.~Matias, and J.~Virto,
\href{http://arxiv.org/abs/1704.05340}{{\tt arXiv:1704.05340 [hep-ph]}}.
%%CITATION = ARXIV:1704.05340;%%.

\bibitem{Geng:2017svp}
L.-S. Geng, B.~Grinstein, S.~JŠger, J.~Martin~Camalich, X.-L. Ren, and R.-X.
  Shi,
\href{http://arxiv.org/abs/1704.05446}{{\tt arXiv:1704.05446 [hep-ph]}}.
%%CITATION = ARXIV:1704.05446;%%.

\bibitem{Ciuchini:2017mik}
M.~Ciuchini, A.~M. Coutinho, M.~Fedele, E.~Franco, A.~Paul, L.~Silvestrini, and
  M.~Valli,
\href{http://arxiv.org/abs/1704.05447}{{\tt arXiv:1704.05447 [hep-ph]}}.
%%CITATION = ARXIV:1704.05447;%%.

\bibitem{DAmico:2017mtc}
G.~D'Amico, M.~Nardecchia, P.~Panci, F.~Sannino, A.~Strumia, R.~Torre, and
  A.~Urbano,
\href{http://arxiv.org/abs/1704.05438}{{\tt arXiv:1704.05438 [hep-ph]}}.
%%CITATION = ARXIV:1704.05438;%%.

\bibitem{Bernlochner:2017jka}
F.~U. Bernlochner, Z.~Ligeti, M.~Papucci, and D.~J. Robinson,
\href{http://arxiv.org/abs/1703.05330}{{\tt arXiv:1703.05330 [hep-ph]}}.
%%CITATION = ARXIV:1703.05330;%%.

\bibitem{Bhattacharya:2014wla}
B.~Bhattacharya, A.~Datta, D.~London, and S.~Shivashankara,
  \href{http://dx.doi.org/10.1016/j.physletb.2015.02.011}{{\em Phys. Lett.}
  {\bf B742} (2015)  370--374},
\href{http://arxiv.org/abs/1412.7164}{{\tt arXiv:1412.7164 [hep-ph]}}.
%%CITATION = ARXIV:1412.7164;%%.

\bibitem{Alonso:2015sja}
R.~Alonso, B.~Grinstein, and J.~Martin~Camalich,
  \href{http://dx.doi.org/10.1007/JHEP10(2015)184}{{\em JHEP} {\bf 10} (2015)
  184},
\href{http://arxiv.org/abs/1505.05164}{{\tt arXiv:1505.05164 [hep-ph]}}.
%%CITATION = ARXIV:1505.05164;%%.

\bibitem{Greljo:2015mma}
A.~Greljo, G.~Isidori, and D.~Marzocca,
  \href{http://dx.doi.org/10.1007/JHEP07(2015)142}{{\em JHEP} {\bf 07} (2015)
  142},
\href{http://arxiv.org/abs/1506.01705}{{\tt arXiv:1506.01705 [hep-ph]}}.
%%CITATION = ARXIV:1506.01705;%%.

\bibitem{Calibbi:2015kma}
L.~Calibbi, A.~Crivellin, and T.~Ota,
  \href{http://dx.doi.org/10.1103/PhysRevLett.115.181801}{{\em Phys. Rev.
  Lett.} {\bf 115} (2015)  181801},
\href{http://arxiv.org/abs/1506.02661}{{\tt arXiv:1506.02661 [hep-ph]}}.
%%CITATION = ARXIV:1506.02661;%%.

\bibitem{Bauer:2015knc}
M.~Bauer and M.~Neubert,
  \href{http://dx.doi.org/10.1103/PhysRevLett.116.141802}{{\em Phys. Rev.
  Lett.} {\bf 116} (2016) no.~14, 141802},
\href{http://arxiv.org/abs/1511.01900}{{\tt arXiv:1511.01900 [hep-ph]}}.
%%CITATION = ARXIV:1511.01900;%%.

\bibitem{Fajfer:2015ycq}
S.~Fajfer and N.~Ko\v{s}nik,
  \href{http://dx.doi.org/10.1016/j.physletb.2016.02.018}{{\em Phys. Lett.}
  {\bf B755} (2016)  270--274},
\href{http://arxiv.org/abs/1511.06024}{{\tt arXiv:1511.06024 [hep-ph]}}.
%%CITATION = ARXIV:1511.06024;%%.

\bibitem{Barbieri:2015yvd}
R.~Barbieri, G.~Isidori, A.~Pattori, and F.~Senia,
  \href{http://dx.doi.org/10.1140/epjc/s10052-016-3905-3}{{\em Eur. Phys. J.}
  {\bf C76} (2016) no.~2, 67},
\href{http://arxiv.org/abs/1512.01560}{{\tt arXiv:1512.01560 [hep-ph]}}.
%%CITATION = ARXIV:1512.01560;%%.

\bibitem{Das:2016vkr}
D.~Das, C.~Hati, G.~Kumar, and N.~Mahajan,
  \href{http://dx.doi.org/10.1103/PhysRevD.94.055034}{{\em Phys. Rev.} {\bf
  D94} (2016)  055034},
\href{http://arxiv.org/abs/1605.06313}{{\tt arXiv:1605.06313 [hep-ph]}}.
%%CITATION = ARXIV:1605.06313;%%.

\bibitem{Boucenna:2016qad}
S.~M. Boucenna, A.~Celis, J.~Fuentes-Martin, A.~Vicente, and J.~Virto,
  \href{http://dx.doi.org/10.1007/JHEP12(2016)059}{{\em JHEP} {\bf 12} (2016)
  059},
\href{http://arxiv.org/abs/1608.01349}{{\tt arXiv:1608.01349 [hep-ph]}}.
%%CITATION = ARXIV:1608.01349;%%.

\bibitem{Becirevic:2016yqi}
D.~Becirevic, S.~Fajfer, N.~Ko\v{s}nik, and O.~Sumensari,
  \href{http://dx.doi.org/10.1103/PhysRevD.94.115021}{{\em Phys. Rev.} {\bf
  D94} (2016) no.~11, 115021},
\href{http://arxiv.org/abs/1608.08501}{{\tt arXiv:1608.08501 [hep-ph]}}.
%%CITATION = ARXIV:1608.08501;%%.

\bibitem{Hiller:2016kry}
G.~Hiller, D.~Loose, and K.~Schoenwald,
  \href{http://dx.doi.org/10.1007/JHEP12(2016)027}{{\em JHEP} {\bf 12} (2016)
  027},
\href{http://arxiv.org/abs/1609.08895}{{\tt arXiv:1609.08895 [hep-ph]}}.
%%CITATION = ARXIV:1609.08895;%%.

\bibitem{Bhattacharya:2016mcc}
B.~Bhattacharya, A.~Datta, J.-P. Gu\'evin, D.~London, and R.~Watanabe,
  \href{http://dx.doi.org/10.1007/JHEP01(2017)015}{{\em JHEP} {\bf 01} (2017)
  015},
\href{http://arxiv.org/abs/1609.09078}{{\tt arXiv:1609.09078 [hep-ph]}}.
%%CITATION = ARXIV:1609.09078;%%.

\bibitem{Buttazzo:2016kid}
D.~Buttazzo, A.~Greljo, G.~Isidori, and D.~Marzocca,
  \href{http://dx.doi.org/10.1007/JHEP08(2016)035}{{\em JHEP} {\bf 08} (2016)
  035},
\href{http://arxiv.org/abs/1604.03940}{{\tt arXiv:1604.03940 [hep-ph]}}.
%%CITATION = ARXIV:1604.03940;%%.

\bibitem{Barbieri:2016las}
R.~Barbieri, C.~W. Murphy, and F.~Senia,
  \href{http://dx.doi.org/10.1140/epjc/s10052-016-4578-7}{{\em Eur. Phys. J.}
  {\bf C77} (2017) no.~1, 8},
\href{http://arxiv.org/abs/1611.04930}{{\tt arXiv:1611.04930 [hep-ph]}}.
%%CITATION = ARXIV:1611.04930;%%.

\bibitem{Bordone:2017anc}
M.~Bordone, G.~Isidori, and S.~Trifinopoulos,
\href{http://arxiv.org/abs/1702.07238}{{\tt arXiv:1702.07238 [hep-ph]}}.
%%CITATION = ARXIV:1702.07238;%%.

\bibitem{Crivellin:2017zlb}
A.~Crivellin, D.~MŸller, and T.~Ota,
\href{http://arxiv.org/abs/1703.09226}{{\tt arXiv:1703.09226 [hep-ph]}}.
%%CITATION = ARXIV:1703.09226;%%.

\bibitem{Glashow:2014iga}
S.~L. Glashow, D.~Guadagnoli, and K.~Lane,
  \href{http://dx.doi.org/10.1103/PhysRevLett.114.091801}{{\em Phys. Rev.
  Lett.} {\bf 114} (2015)  091801},
\href{http://arxiv.org/abs/1411.0565}{{\tt arXiv:1411.0565 [hep-ph]}}.
%%CITATION = ARXIV:1411.0565;%%.

\bibitem{Barbieri:2011ci}
R.~Barbieri, G.~Isidori, J.~Jones-Perez, P.~Lodone, and D.~M. Straub,
  \href{http://dx.doi.org/10.1140/epjc/s10052-011-1725-z}{{\em Eur. Phys. J.}
  {\bf C71} (2011)  1725},
\href{http://arxiv.org/abs/1105.2296}{{\tt arXiv:1105.2296 [hep-ph]}}.
%%CITATION = ARXIV:1105.2296;%%.

\bibitem{Barbieri:2012uh}
R.~Barbieri, D.~Buttazzo, F.~Sala, and D.~M. Straub,
  \href{http://dx.doi.org/10.1007/JHEP07(2012)181}{{\em JHEP} {\bf 07} (2012)
  181},
\href{http://arxiv.org/abs/1203.4218}{{\tt arXiv:1203.4218 [hep-ph]}}.
%%CITATION = ARXIV:1203.4218;%%.

\bibitem{Buchalla:1993wq}
G.~Buchalla and A.~J. Buras,
  \href{http://dx.doi.org/10.1016/0550-3213(94)90496-0}{{\em Nucl. Phys.} {\bf
  B412} (1994)  106--142},
\href{http://arxiv.org/abs/hep-ph/9308272}{{\tt arXiv:hep-ph/9308272
  [hep-ph]}}.
%%CITATION = HEP-PH/9308272;%%.

\bibitem{Mescia:2007kn}
F.~Mescia and C.~Smith,
  \href{http://dx.doi.org/10.1103/PhysRevD.76.034017}{{\em Phys. Rev.} {\bf
  D76} (2007)  034017},
\href{http://arxiv.org/abs/0705.2025}{{\tt arXiv:0705.2025 [hep-ph]}}.
%%CITATION = ARXIV:0705.2025;%%.

\bibitem{Isidori:2005xm}
G.~Isidori, F.~Mescia, and C.~Smith,
  \href{http://dx.doi.org/10.1016/j.nuclphysb.2005.04.008}{{\em Nucl. Phys.}
  {\bf B718} (2005)  319--338},
\href{http://arxiv.org/abs/hep-ph/0503107}{{\tt arXiv:hep-ph/0503107
  [hep-ph]}}.
%%CITATION = HEP-PH/0503107;%%.

\bibitem{Buchalla:1998ba}
G.~Buchalla and A.~J. Buras,
  \href{http://dx.doi.org/10.1016/S0550-3213(99)00149-2}{{\em Nucl. Phys.} {\bf
  B548} (1999)  309--327},
\href{http://arxiv.org/abs/hep-ph/9901288}{{\tt arXiv:hep-ph/9901288
  [hep-ph]}}.
%%CITATION = HEP-PH/9901288;%%.

\bibitem{Buras:2006gb}
A.~J. Buras, M.~Gorbahn, U.~Haisch, and U.~Nierste,
  \href{http://dx.doi.org/10.1007/JHEP11(2012)167,
  10.1088/1126-6708/2006/11/002}{{\em JHEP} {\bf 11} (2006)  002},
  \href{http://arxiv.org/abs/hep-ph/0603079}{{\tt arXiv:hep-ph/0603079
  [hep-ph]}}.
[Erratum: JHEP11,167(2012)].
%%CITATION = HEP-PH/0603079;%%.

\bibitem{Buras:2015qea}
A.~J. Buras, D.~Buttazzo, J.~Girrbach-Noe, and R.~Knegjens,
  \href{http://dx.doi.org/10.1007/JHEP11(2015)033}{{\em JHEP} {\bf 11} (2015)
  033},
\href{http://arxiv.org/abs/1503.02693}{{\tt arXiv:1503.02693 [hep-ph]}}.
%%CITATION = ARXIV:1503.02693;%%.

\bibitem{Olive:2016xmw}
{\bf Particle Data Group}, C.~Patrignani {\em et al.,}
\href{http://dx.doi.org/10.1088/1674-1137/40/10/100001}{{\em Chin. Phys.} {\bf
  C40} (2016) no.~10, 100001}.
%%CITATION = CHPHD,C40,100001;%%.

\bibitem{Ruggiero:2017hjh}
{\bf NA62}, G.~Ruggiero
\href{http://dx.doi.org/10.1088/1742-6596/800/1/012023}{{\em J. Phys. Conf.
  Ser.} {\bf 800} (2017) no.~1, 012023}.
%%CITATION = 00462,800,012023;%%.

\bibitem{Beckford:2017qze}
{\bf KOTO}, B.~Beckford
{\em PoS} {\bf ICHEP2016} (2017)  580.
%%CITATION = POSCI,ICHEP2016,580;%%.

\bibitem{Faroughy:2016osc}
D.~A. Faroughy, A.~Greljo, and J.~F. Kamenik,
  \href{http://dx.doi.org/10.1016/j.physletb.2016.11.011}{{\em Phys. Lett.}
  {\bf B764} (2017)  126--134},
\href{http://arxiv.org/abs/1609.07138}{{\tt arXiv:1609.07138 [hep-ph]}}.
%%CITATION = ARXIV:1609.07138;%%.

\bibitem{Feruglio:2017rjo}
F.~Feruglio, P.~Paradisi, and A.~Pattori,
\href{http://arxiv.org/abs/1705.00929}{{\tt arXiv:1705.00929 [hep-ph]}}.
%%CITATION = ARXIV:1705.00929;%%.

\bibitem{Feruglio:2016gvd}
F.~Feruglio, P.~Paradisi, and A.~Pattori,
  \href{http://dx.doi.org/10.1103/PhysRevLett.118.011801}{{\em Phys. Rev.
  Lett.} {\bf 118} (2017) no.~1, 011801},
\href{http://arxiv.org/abs/1606.00524}{{\tt arXiv:1606.00524 [hep-ph]}}.
%%CITATION = ARXIV:1606.00524;%%.

\bibitem{Buras:2014fpa}
A.~J. Buras, J.~Girrbach-Noe, C.~Niehoff, and D.~M. Straub,
  \href{http://dx.doi.org/10.1007/JHEP02(2015)184}{{\em JHEP} {\bf 02} (2015)
  184},
\href{http://arxiv.org/abs/1409.4557}{{\tt arXiv:1409.4557 [hep-ph]}}.
%%CITATION = ARXIV:1409.4557;%%.

\bibitem{Zupanc:2017CERN}
{\bf Belle II}, A.~Zupanc.
  \href{https://indico.cern.ch/event/633880/contributions/2577392/attachments/1462862/2260081/BelleII_Zupanc.pdf}{Talk
  at Instant Workshop on B meson anomalies, CERN, 2017}.

\end{thebibliography}\endgroup
}

\end{document}